\documentclass[twocolumn,english,prl,floatfix,showpacs]{revtex4}
\usepackage[T1]{fontenc}
\usepackage[latin9]{inputenc}

\usepackage{babel}
\usepackage{graphicx}
\usepackage{subfig}

\begin{document}

\title{An efficient numerical algorithm on irreducible multiparty correlations}

\author{D.L. Zhou}

\affiliation{Beijing National Laboratory for Condensed Matter Physics, and Institute of Physics, Chinese Academy of Sciences, Beijing 100190, China}

\begin{abstract}
We develop a numerical algorithm to calculate the degrees of irreducible multiparty correlations for an arbitrary multiparty quantum state, which is efficient for any quantum state of up to five qubits. We demonstrate the power of the algorithm by the explicit calculations of the degrees of irreducible multiparty correlations in the $4$-qubit GHZ state, the Smolin state, and the $5$-qubit W state. This development takes a crucial step towards practical applications of irreducible multiparty correlations in real quantum many-body physics.
\end{abstract}

\pacs{03.67.Mn, 03.65.Ud, 89.70.Cf}

\maketitle

\textit{Introduction}.---
In modern physics, as is well known, mean field theory is not
sufficient to describe the physics in a strongly correlated
many-body system \cite{WWF06}, which implies that there exist rich correlation
structures in its quantum state. Therefore, how to characterize
multiparty correlations in a multipartite quantum state becomes a
fundamental problem in many-body physics. Traditional physical
method is to introduce the correlation functions to describe the
correlations in a many-body system.

The extensive researches on characterizing entanglement in quantum
information science \cite{HHHH09} shed different light on the problem. On one
hand, correlation functions are not invariant under local unitary
transformations, which implies that they can be regarded only as
correlation witnesses but not as legitimate correlation measures \cite{HV01,ZZXY06}. On
the other hand, the information based viewpoint can be instructive
in characterizing correlations in a multipartite quantum state.

 In the information viewpoint, the degree of the  total correlation \cite{Wat60}  in a multipartite quantum system is equal to the difference between the sum of von Neumann entropies of all the subsystems and the von Neumann entropy of the whole system.  There are two different schemes to classify the total correlation:  one is to distinguish the total correlation into quantum correlation and classical correlation \cite{HV01,GPW05},  the other is to divide the total correlation into pairwise correlation, triplewise correlation, etc..
 
 The concept of irreducible $n$-party correlation in an $n$-partite quantum state was first proposed in Ref. \cite{LPW02}. We generalized it to irreducible $m$-party ($2\le m \le N$) correlation in an $n$-partite state, and proposed that all the irreducible $m$-party correlations construct a classification of the total correlation \cite{Zho08,Zho09}. It is worthy to note that, in classical information community, the irreducible $m$-party correlations in a joint probability distribution of $n$ classical random variables  were investigated in Ref. \cite{Ama01,SSBB03}.

The degree of irreducible multiparty correlations in a multipartite quantum state, like many important quantities in quantum information science, such as
the measure of entanglement \cite{BDS+96} and the capacity of a
quantum channel \cite{Hol98,SW97}, are defined as an optimization
problem, which makes their calculations become extremely difficult.
These computational difficulties almost prevent any practical application
of these measures in a real physical problem. Therefore it is of
great significance to develop an efficient algorithm to calculate
them for a general multiparty quantum state.

In Ref. \cite{Zho08}, we proposed a continuity approach that reduces
the calculations of irreducible multiparty correlations in a multiparty
quantum state without maximal rank to the calculations of irreducible
multiparty correlations in a series of multiparty quantum states with
maximal rank. Although theorem 1 in Ref. \cite{Zho08} tells us the
form of a maximal rank state without higher order irreducible multiparty
correlations, however, this theorem does not solve the problem on
the calculations of irreducible multiparty correlations for a general
multiparty quantum state with maximal rank. This is why we solve the
calculations only for some specific classes of states in Ref. \cite{Zho08}.
In other words, we have not a systematic method to calculate the irreducible
multiparty correlations for a state with maximal rank.

In this Letter, we  develop an efficient systematic numerical
algorithm on the calculations of the degrees of irreducible
multiparty correlations for a general multipartite quantum state.The advantages of our algorithm is that it is
independent of initial values of variables, and we find it is
efficient for an arbitrary quantum state of up
to five qubits. To the best of our knowledge, it is for the first time that we have the capacity to
deal with the detailed analysis of the correlations in a general multiparty state of up to five qubits.

\textit{Notations and Definitions}.---
The Hilbert space of an $n$-partite quantum system is denoted by $\mathcal{H}^{[n]}\equiv\prod_{i=1}^{n}\otimes\mathcal{H}^{(i)}$,
where $[n]$ is the set $\{1,2,\cdots,n\}$, and $\mathcal{H}^{(i)}$ is the Hilbert space of party $i$ whose
dimension is $d_{i}$. The inner product of two operators $A^{(i)}$
and $B^{(i)}$ in the Hilbert space $\mathcal{H}^{(i)}$ is defined
as $\langle A^{(i)}\vert B^{(i)}\rangle=\frac{1}{d_{i}}\mathrm{Tr}(A^{(i)\dagger}B^{(i)})$ \cite{HR85}. The prefactor $\frac {1} {d_{i}}$ is introduced to satisfy the normalization condition $\langle I^{(i)}\vert I^{(i)}\rangle=1$, where $I^{(i)}$ is the identity operator in the Hilbert space $\mathcal{H}_{i}$.
Thus we can introduce an orthonormal Hermitian operator basis $\{O_{a_{i}}^{(i)},a_{i}\in\{0,1,\cdots,d_{i}^{2}-1\}\}$.
In particular, we take $O_{0}^{(i)}$ to be the identity operator $I^{(i)}$.
Any operator $A^{(i)}$ can be expanded in this basis as $A^{(i)}=\sum_{a_{i}}O_{a_{i}}^{(i)}\langle O_{a_{i}}^{(i)}\vert A^{(i)}\rangle$.
Further more, the operator $A^{[n]}$ in the $n$-party Hilbert space
can be expanded as $A^{[n]}=\sum_{\mathbf{a}(n)}O_{\mathbf{a}(n)}^{[n]}\langle O_{\mathbf{a}(n)}^{[n]}\vert A^{[n]}\rangle$, where $\mathbf{a}(n)$ is the set $\{a_{1},a_{2},\cdots,a_{n}\}$, and $O^{[n]}_{\mathbf{a}(n)}$ is an abbreviated notation for $\prod_{i=1}^{n}O^{(i)}_{a_{i}}$.
If the operator $A^{[n]}$ is the Hamiltonian of an $n$-party system,
the terms to describe $m$-party interactions $(1\le m\le n)$ satisfy
the condition $N_{0}(\mathbf{a}(n))=n-m$ with $N_{0}(\mathbf{a}(n))=\sum_{a_{i}}\delta_{0a_{i}}$.

Without loss of generality, we consider an $n$-party quantum state
$\rho^{[n]}$ with maximal rank, which can be expanded as
\begin{equation}
\rho^{[n]}=\sum_{\mathbf{a}(n)}O_{\mathbf{a}(n)}^{[n]}\langle O_{\mathbf{a}(n)}^{[n]}\vert\rho^{[n]}\rangle.\label{eq:rhon}
\end{equation}
Because the state $\rho^{[n]}$ is positive definite, we can define
$\ln\rho^{[n]}$ uniquely as a Hermitian operator. Then we can apply
the above expansion to $\ln\rho^{[n]}$ to obtain
\begin{equation}
\ln\rho^{[n]}=\sum_{\mathbf{a}(n)}O_{\mathbf{a}(n)}^{[n]}\langle O_{\mathbf{a}(n)}^{[n]}\vert\ln\rho^{[n]}\rangle.\label{eq:lnrhon}
\end{equation}
The condition $\mathrm{Tr}\rho^{[n]}=1$ implies that the coefficient
$\langle O_{\mathbf{0}(n)}^{[n]}\vert\ln\rho^{[n]}\rangle$ can be
determined by the other coefficients $\langle O_{\mathbf{\tilde{a}}(n)}^{[n]}\vert\ln\rho^{[n]}\rangle$. Here $\mathbf{0}(n)$ is the set $\mathbf{a}(n)$ with $a_{i}=0$ for $i\in [n]$, and $\mathbf{\tilde{a}}(n)$ is the same as $\mathbf{a}(n)$ except $\mathbf{0}(n)$.
Compared with the expansion (\ref{eq:rhon}), the obvious advantage
of the expansion (\ref{eq:lnrhon}) is that it ensures the positivity
of $\rho^{[n]}$ automatically. Further more, a one-to-one map between
the state $\rho^{[n]}$ with maximal rank and the set of real coefficients
$\{\langle O_{\mathbf{\tilde{a}}(n)}^{[n]}\vert\ln\rho^{[n]}\rangle\}$
can be built. The existence of such a one-to-one map is an essential
element in our numerical algorithm.

To make use of the expansion (\ref{eq:lnrhon}), we adopt the equivalent
definitions of the degrees of irreducible multiparty correlations
in a multiparty quantum state given in Ref. \cite{Zho09} but not
the original definitions proposed in Ref. \cite{LPW02,Zho08}. If
we adopt the original definition, then the optimization is made under
the expansion (\ref{eq:rhon}) , which makes the optimization almost
impossible because of the constraint of semipositivity of a density matrix. In Ref. \cite{Zho09}, we give the definitions on the
degrees of irreducible multiparty correlations for a three-qubit system.
Now the definitions are generalized for a general multipartite quantum state
with a finite dimensional Hilbert space as follows.

We first define the set of the $n$-party states without more-than-$m$-party
irreducible correlations as \begin{equation}
B_{m}\equiv\{\sigma^{[n]}\vert\langle O_{\mathbf{a}(n)}^{[n]}\vert\ln\sigma^{[n]}\rangle=0,\forall N_{0}(\mathbf{a}(n))<n-m\}.\label{eq:Bm}\end{equation}
Next we find the state in the set $B_{m}$ that is least distinguishable
with the state $\rho^{[n]}$ \begin{equation}
\rho_{m}^{[n]}\equiv\arg\min_{\sigma^{[n]}\in B_{m}}S(\rho^{[n]}\vert\vert\sigma^{[n]}),\label{eq:rhonm}\end{equation}
where the quantum relative entropy \cite{Ved02} $S(\rho\vert\vert\rho^{\prime})=\mathrm{Tr}(\rho(\ln\rho-\ln \rho^{\prime}))$ for two quantum states $\rho$ and $\rho^{\prime}$ in the same Hilbert space.
Then the degree of irreducible $m$-party correlation is defined as
\begin{equation}
C_{m}(\rho^{[n]})\equiv S(\rho_{m}^{[n]}\vert\vert\rho_{m-1}^{[n]}).\label{eq:cm}
\end{equation}
In addition, the degree of the total correlation is defined by
\begin{equation}
C_{T}(\rho^{[n]})\equiv S(\rho^{[n]}\vert\vert\rho_{1}^{[n]}).\label{eq:ct}
\end{equation}
Using the same arguments given in Ref. \cite{Zho09}, we can show
that $C_{T}(\rho^{[n]})=\sum_{m=2}^{n}C_{m}(\rho^{[n]})=\sum_{i=1}^{n}S(\rho^{(i)})-S(\rho^{[n]})$ with the von Neaumann entropy $S(\rho)=-\mathrm{Tr}(\rho\ln\rho)$ for a quantum state $\rho$.

\textit{Numerical Algorithm}.---
In the above optimization problem, it is an essential task to find out the
state $\rho_{m}^{[n]}$ for a given state $\rho^{[n]}$. It is possible
to directly solve Eq. (\ref{eq:rhonm}) to obtain the state $\rho_{m}^{[n]}$.
However, it is doutful whether the solution we find is a local minimum or a global minimum.
Fortunately, the optimization problem (\ref{eq:rhonm}) can be transformed
into the following system of nonlinear equations:
\begin{eqnarray}
&&\langle O_{\mathbf{a}(n)}^{[n]}\vert\ln\rho_{m}^{[n]}\rangle  =  0,\;\;\;\;\forall N_{0}(\mathbf{a}(n))<n-m,\label{eq:rhonma}\\
&&\langle O_{\mathbf{a}(n)}^{[n]}\vert\rho_{m}^{[n]}\rangle  = \langle O_{\mathbf{a}(n)}^{[n]}\vert\rho^{[n]}\rangle,\;\;\;\forall N_{0}(\mathbf{a}(n))\ge m.\label{eq:rhonmb}
\end{eqnarray}

In Ref. \cite{Zho09}, we proved that there exists a unique real solution of
$\{\langle O_{\mathbf{\tilde{a}}(n)}^{[n]}\vert\ln\rho_{m}^{[n]}\rangle\}$
satisfying the above system of equations for a three-qubit system. This result is also valid for a general multipartite quantum state with a finite dimensional Hilbert space.  Here we neglect the proof because it is a simple generalization for the three-qubit case. Thus we have two different ways
to use the system of
equations (\ref{eq:rhonma}, \ref{eq:rhonmb}). On one hand, we can use them to verify
whether the solution of the optimization problem (\ref{eq:rhonm})
is correct. On the other hand, we can directly use the optimization
method to solve them  to obtain the states $\rho_{m}^{[n]}$. In our present numerical algorithm, we adopt the latter method in application of the system of equations  (\ref{eq:rhonma}, \ref{eq:rhonmb}). We want to emphasize that Eq. (\ref{eq:lnrhon}) is used to represent a multipartite quantum state in our algorithm.

For an optimization problem, one of the key skills is to choose a proper initial
value. Here we adopt a continuity approach to choose a proper initial
value for any $n$-partite quantum state $\rho^{[n]}$. We consider a seriers
of states 
\begin{equation}
\rho^{[n]}(p_{0})=p_{0}\frac{I^{[n]}}{d^{[n]}}+(1-p_{0})\rho^{[n]}.\label{eq:rhonp0}
\end{equation}
We take $p_{0}=1-\frac{k}{N}$ with $k\in\{0,\cdots,N\}$, where $N$
is a large positive integer. Obviously, $\rho^{[n]}(k=0)=\frac{I^{[n]}}{d^{[n]}}$
, $\rho^{[n]}(k=N)=\rho^{[n]}$, and $\langle O_{\mathbf{\tilde{a}}(n)}^{[n]}\vert\ln\rho_{m}^{[n]}(k=0)\rangle=0$.
We take the values of $\{\langle O_{\mathbf{\tilde{a}}(n)}^{[n]}\vert\ln\rho_{m}^{[n]}(k)\rangle\}$
as the initial values of  $\{\langle O_{\mathbf{\tilde{a}}(n)}^{[n]}\vert\ln\rho_{m}^{[n]}(k+1)\rangle\}$ for $k=0,1,\cdots,N-1$.

The basic idea under the above approach is based on the continuity
principle, more precisely, the state $\rho^{[n]}(k+1)$ is very similar
to the state $\rho^{[n]}(k)$, so the values of $\{\langle O_{\mathbf{\tilde{a}}(n)}^{[n]}\vert\ln\rho_{m}^{[n]}(k+1)\rangle\}$
is also near the values $\{\langle O_{\mathbf{\tilde{a}}(n)}^{[n]}\vert\ln\rho_{m}^{[n]}(k)\rangle\}$.
The practice of our computations shows that our selection of initial values makes the algorithm become
efficient. The cost of the algorithm is that we calculate the degrees
of irreducible multiparty correlations for a series of states $\rho^{[n]}(p_{0})$
instead of a single state $\rho^{[n]}$.

An obvious advantage is that the choose of initial values in our algorithm
is independent of the state $\rho^{[n]}$. In other words, our algorithm
makes the computation of a general multiparty state become efficient.
In my personal computer, it is efficient for any state up to five qubits. To the best of my knowledge,
it is the best results on multiparty correlations in a multiparty state we obtained so far.

\textit{Numerical results}.---
We will demonstrate the power of our numerical algorithm by explicitly giving the results on irreducible multiparty correlations for some typical multiparty states: the $4$-qubit GHZ state \cite{GHZ89}, the four-qubit Smolin state \cite{Smo01}, and the $5$-qubit W state \cite{DVC00}.

The first state we consider is the $n$-qubit GHZ state $|\textrm{GHZ}_{n}\rangle=\frac {1} {\sqrt{2}} (\prod_{i=1}^{n} \otimes|0\rangle_{i}+\prod_{i=1}^{n}\otimes|1\rangle_{i})$. The degrees of irreducible multiparty correlations on the $4$-qubit GHZ state are given in Figure 1. The total correlation in the state is $4$ bits, and it is classified into $3$ bits of irreducible two-qubit correlation and $1$ bit of irreducible four-qubit correlation. These results are the same as those given in Ref. \cite{Zho08}, and they are consistent with the conclusion in Ref. \cite{WL08, LW02}.

\begin{figure}[htbp]
\begin{center}
\centering
\includegraphics[width=0.5\textwidth]{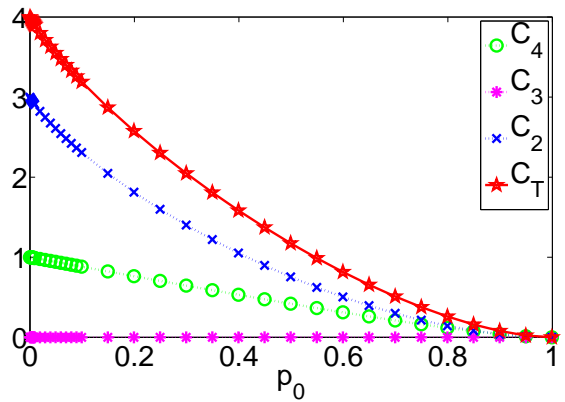}
\caption{The degrees of irreducible multiparty correlations for the $4$-qubit GHZ state.}
\end{center}
\end{figure}

The second state we consider is the $4$-qubit Smolin state, whose density matrix is simply given by $\rho^{[4]}_{\textrm{smo}}=\frac {1} {16} (I^{[4]}+\prod_{i=1}^{4}\sigma_{x}^{(i)}+\prod_{i=1}^{4} \sigma_{z}^{(i)})$. We find that there exists $2$ bits of correlation in the state, and they are irreducible $4$-qubit correlations, which is shown in Figure 2. From the density matrix of the Smolin state, we know that it is also a generalized stabilizer state defined in Ref. \cite{Zho08}. In this sense, the numerical results also verify the results in Ref. \cite{Zho08}.

\begin{figure}[htbp]
\begin{center}
\centering
\includegraphics[width=0.5\textwidth]{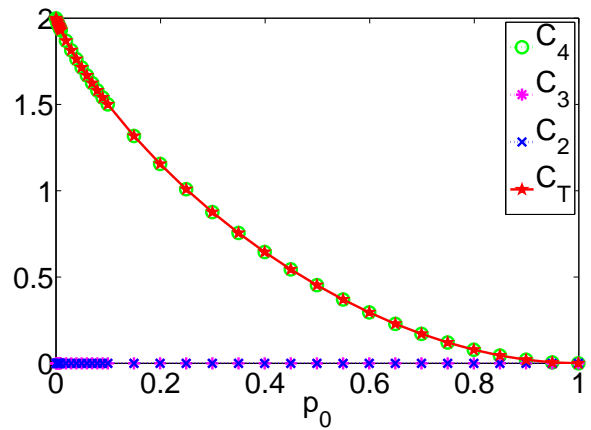}
\caption{The degrees of irreducible multiparty correlations for the four-qubit Smolin state}.
\end{center}
\end{figure}

The third state we consider is the $5$-qubit W state $\vert \textrm{W}_{5}\rangle=\frac {1} {\sqrt{n}} \sum_{i=1}^{n} \sigma_{x}^{(i)}\prod_{j=1}^{5} \otimes\vert 0\rangle_{j}$. Our numerical results show that only irreducible two-qubit correlations exist in the W state, which support the conclusion in Ref. (\cite{PR09}). 
\begin{figure}[htbp]
\begin{center}
\includegraphics[width=0.5\textwidth]{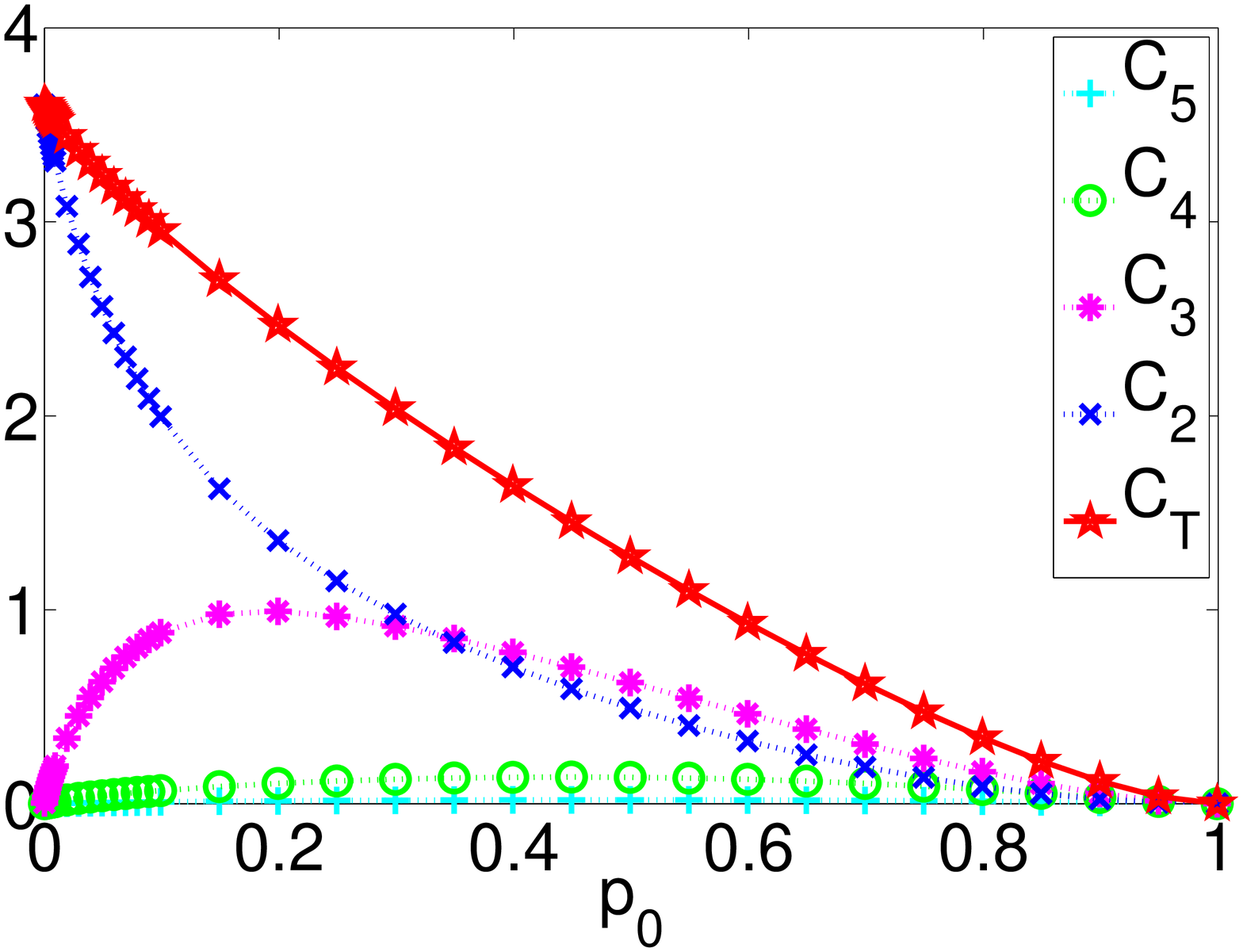}
\caption{The degrees of irredcible multiparty correlations of the $5$-qubit W state. }
\end{center}
\end{figure}

In the range of our numerical results, we find that the degree of the total correlation $C_{T}$ is a non-increasing function of $p_{0}$, however, the degree of irreducible $m$-party correlation can increase with increasing $p_{0}$ (see, for example, Figure 3). Actually we can prove that $C_{T}(\rho^{[n]}(p_{0}))$ is a non-increasing function of $p_{0}$ for any $n$-party state $\rho^{[n]}$ as follows.  We can imagine that every subsystems of the $n$-partite quantum system pass through a depolarized channel \cite{NC00}, then the  quantum state $\rho^{[n]}$ evolves according to Eq. (\ref{eq:rhonp0}) in the direction of increasing $p_0$. In the process, only local operations act on the state $\rho^{[n]}$, and the degree of total correlation does not increase under local operations, therefore  $C_{T}(\rho^{[n]}(p_{0}))$ is a non-increasing function of $p_{0}$ for any $n$-party state $\rho^{[n]}$. 
 
In addition, the fact $C_{m}$ ($m=3, 4, 5$) is not a non-increasing for a $5$-qubit W state gives another example to support one of the main results in Ref. \cite{Zho09}: local operations can transform lower order correlations into higher order correlations.

\textit{Discussions and summary}.---
The calculations of the degrees of irreducible multiparty correlations
for an arbitrary multiparty quantum state are challenging because
they are defined as the constraint optimization problems over all the multiparty
quantum states in the whole Hilbert space.  In this Letter, we develop an efficient numerical method to calculate the degrees of irreducible multiparty correlations for any multipartite quantum state, which is based on the following two key elements.

One key element in our algorithm is that we adopt the expansion of a multipartite state in the exponential form (\ref{eq:lnrhon}). First, it ensures the positivity of the state automatically. Second, although the independent varibles $\{ \langle O_{\mathbf{\tilde{a}}(n)}^{[n]}\vert\ln\rho^{[n]}\rangle\}$ can take the limit to infinity, the state $\rho^{[n]}$ is always well defined because of the constraint $\mathrm{Tr}\rho^{[n]}=1$. In this sense, the state without maximal rank are naturally contained in this expansion if the coefficients can limit to infinity. This makes our algorithm effective for arbitrary multiparty states.

The other key element is related to the selection of the initial values of variables, more precisely, the formula (\ref{eq:rhonp0}). It makes our algorithm independent on the initial values of variables, and greatly enhances the efficiency of our algorithm.

In summary, we present an efficient numerical algorithm on the calculations of the degrees of irreducible multiparty correlations in a multiparty quantum state. Our algorithm is valid for arbitrary quantum states up to five qubits in my personal computer, and it is a universal algorithm whose efficiency does not depend strongly on the multipartite quantum state. We demonstrate the power of our algorithm by explicitly giving the results for the $4$-qubit GHZ state, the Smolin state, and the $5$-qubit W state, which are consistent with previous results. We hope that our development of this algorithm will provide a powerful tool to analyze the correlation distributions in a multipartite quantum state, and thus takes a crucial step towards the practical applications of irreducible multiparty correlations in real quantum many-body systems.

The author thank Dr. S. Yang and Prof. C.P. Sun for stimulating discussions. This work is supported
by NSF of China under Grant No. 10775176, and NKBRSF of China under
Grants No. 2006CB921206 and No. 2006AA06Z104.

\end{document}